\documentclass[12pt]{article}

\usepackage{sbc-template}
\usepackage{graphicx,url}
\usepackage[utf8, latin1]{inputenc}
\usepackage{wrapfig}

\title{Towards Polyglot Data Processing in Social Networks\\ using the Hadoop-Spark ecosystem}

\author{Antony Seabra de Medeiros, Sergio Lifschitz}

\address{Departamento de Informatica - PUC-Rio
  \email{\{amedeiros, sergio\}@inf.puc-rio.br}
}

\begin{document} 
\maketitle
\begin{abstract}
This article explores the use of the Hadoop-Spark ecosystem for social media data processing, adopting a polyglot approach with the integration of various computation and storage technologies, such as Hive, HBase and GraphX. We discuss specific tasks involved in processing social network data, such as calculating user influence, counting the most frequent terms in messages and identifying social relationships among users and groups. We conducted a series of empirical performance assessments, focusing on executing selected tasks and measuring their execution time within the Hadoop-Spark cluster. These insights offer a detailed quantitative analysis of the performance efficiency of the ecosystem tools. We conclude by highlighting the potential of the Hadoop-Spark ecosystem tools for advancing research in social networks and related fields.
\end{abstract}

\section{Introduction}
The issue of processing social media data is particularly relevant to businesses and organizations that use social media as a key part of their marketing and customer engagement strategies. These businesses need to understand how their social media activities are impacting their brand, customer satisfaction, and overall business performance. Data processing is a critical part of achieving those insights. Additionally, social media data processing is relevant to researchers and academics who study social media behavior and its impact on society. This includes researchers in fields such as sociology, psychology, political science, and communication studies, who may use social media data to explore topics such as social movements, public opinion, and online behavior.

Over the past two decades, the Hadoop ecosystem has evolved significantly from its initial use as a solution for distributed web page indexing to its current role as a comprehensive platform. This transformation has seen Hadoop become a foundational system for constructing data lakes, adept at handling structured, semi-structured, and unstructured data. The ecosystem initially relied on technologies like MapReduce and Yarn. However, with the advent of the Hadoop-Spark ecosystem, it addressed performance issues encountered in large-scale data processing tasks, particularly prevalent in social media contexts. Spark, integrated into this ecosystem, enhanced processing capabilities, especially for handling voluminous and complex data sets. Furthermore, the development of various tools, including Hive and HBase, has augmented the ecosystem's ability to efficiently manage and process queries from diverse sources.

In the context of social network data analysis, particularly from platforms like Instagram, Facebook, and Twitter, it's crucial to match the processing mechanism with the specific type of query or analysis required. For example, when the goal is to identify frequently used terms associated with a certain entity, utilizing MapReduce or Spark transformations for count jobs can be highly effective. Conversely, for tasks like determining the most influential figures in relation to a specific topic, employing a graph database job could be more advantageous. These challenges can be addressed by a polyglot persistence approach by combining the benefits of several data stores and their underlying techniques \cite{sadalage2013nosql}. Additionally, complementing this with a polyglot computing approach allows for the tailoring of processing capabilities to match the most appropriate technology for each specific need, optimizing performance and efficiency.

We argue that Polyglot Data Processing, a mixture of polyglot computing and polyglot persistence, is a versatile approach in data analysis that uses multiple processing engines and data stores to efficiently handle different types of data and processing needs. In environments like Hadoop and Spark, it allows for the use of specialized tools such as Hive for SQL-like querying, GraphX for graph computations, and MapReduce for general large-scale data processing. This approach is particularly beneficial for complex tasks like social network analysis, as it leverages the strengths of each tool for more insightful, efficient, and scalable data processing, allowing for the optimization of data processing pipelines.

The aim of this study is to discern the present components of the Hadoop-Spark ecosystem that can be used in conjunction to process social network data in a polyglot data processing approach. By confronting the tasks related to social networks with the specificities of the ecosystem components, we can conclude about which of them are more suitable for one or another task.

The paper is structured as follows. Section 2 presents a technical background on Hadoop and Spark technologies. Related work is presented in Section 3. Our methodology is outlined in Section 4. In Section 5, we evaluate the polyglot data processing approach through the execution of selected tasks and, in Section 6, we derive some conclusions from our work and suggest future research directions in this field.

\section{Background}
Hadoop is an open-source platform for distributed storage and processing of large-scale datasets, inspired by the Google File System (GFS) \cite{ghemawat2003google} and MapReduce \cite{dean2004mapreduce} papers published by Google. Hadoop was created by Doug Cutting and Mike Cafarella in 2005, while they were working at Yahoo. The name Hadoop comes from a toy elephant owned by Cutting's son, which also served as the inspiration for the project's logo.

Initially, Hadoop consisted of two main components: Hadoop Distributed File System (HDFS) for distributed storage, and MapReduce for distributed processing. Hadoop was designed to address the challenges of processing and analyzing large-scale datasets that were too big to fit on a single machine's storage or memory.

Hadoop quickly gained popularity in the early 2000s due to its ability to process and analyze massive datasets, making it possible to perform tasks that were previously impossible or extremely time-consuming. In addition, Hadoop's open-source nature allowed for widespread adoption, leading to a vibrant and active community of developers contributing to its ongoing development and evolution.

Over the years, Hadoop has evolved significantly, with the addition of new components, such as Yarn, which became the default cluster management tool in Hadoop 2.0, and a range of complementary tools and frameworks. The Hadoop ecosystem has become a critical part of the big data technology stack, enabling organizations to store, process and conduct data-oriented analysis of large-scale datasets that facilitate the generation of insights and the realization of business value.

\subsection{HDFS}
HDFS is a distributed file system that provides a solution to the problem of storing data across multiple machines in a cluster. Its design has a master/slave architecture with a single Name Node as the master server which manages the file system namespace and regulates access to files by clients. The slaves are a number of Data Nodes, usually one per node in the cluster, which manage storage attached to the nodes that they run on \cite{karun2013review}. The system is designed to ensure fault tolerance, scalability, and efficient storage and retrieval of large-scale datasets.

In HDFS, data is stored in the form of blocks or chunks, which are typically large and set to a default size of 128 MB in Hadoop 2.x and 3.x. By default, each block is replicated three times, though this configuration can be customized on a per-file basis. These replicas are then distributed across nodes in the Hadoop cluster, which ensures both fault tolerance and efficient processing of data. The HDFS NameNode keeps track of the location of these chunks and is responsible for managing file system namespace operations. When a client application requests location information from the Namenode, it responds with the relevant chunk handle and chunk locations. If a certain location is unavailable, the client automatically selects the next available location and retries the request. This behavior is critical in maintaining fault tolerance in HDFS.

\subsection{MapReduce}
MapReduce is a programming model and software framework that enables the distributed processing of large data sets on clusters of computers. As the large datasets are divided into smaller pieces in HDFS, tasks can then be processed in parallel on the different Data Nodes.

MapReduce consists of two primary operations: map and reduce. The map operation takes a set of input data and produces an intermediate set of key-value pairs. The reduce operation then takes these intermediate key-value pairs and combines them to produce a final set of output values. The map and reduce operations are both designed to be highly parallelizable, which makes MapReduce well-suited for distributed computing environments.

\begin{figure}[ht]
\centering
\includegraphics[width=.60\textwidth]{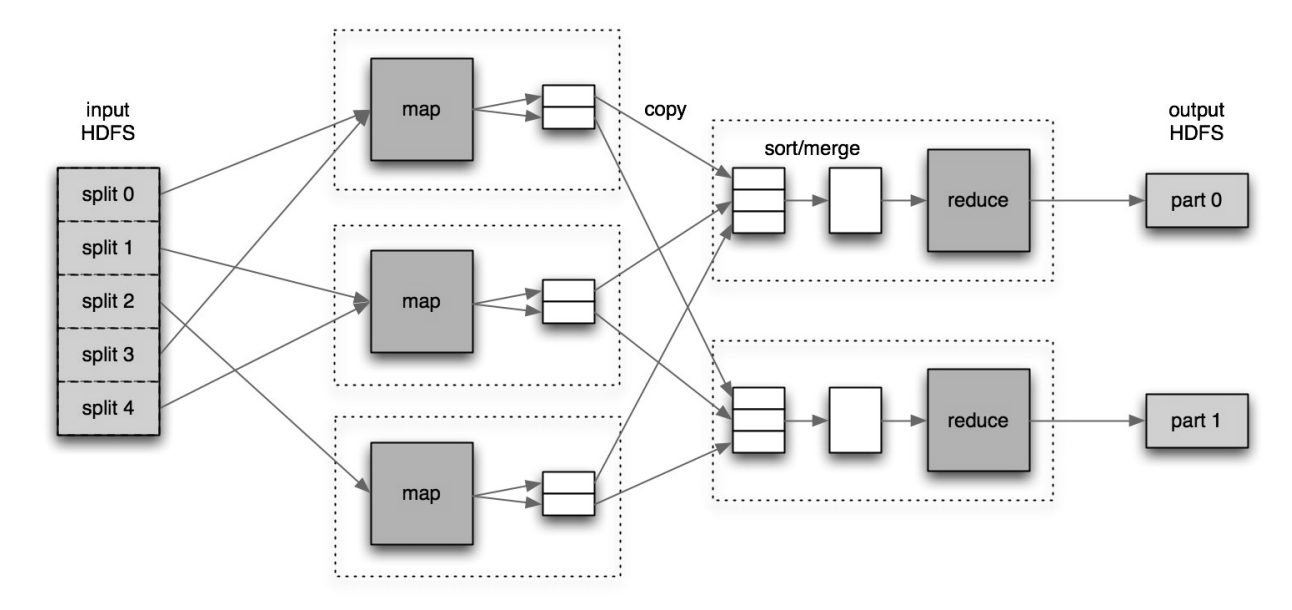}
\caption{Understanting MapReduce with Hadoop} \cite{mapreduce}
\end{figure}

As illustrated in Figure 1, the input comprises three lines which are partitioned into three parts during the Splitting phase and forwarded to individual Map-type tasks. During the Mapping phase, each Map task operates on its respective input on a cluster node and generates a key-value structure. In the specific example, the Map task returns the word count. Subsequently, in the Shuffling phase, data is transferred from the Mappers nodes to the Reducers nodes. Prior to the transfer, a sort is performed on all the keys, which ensures that the same keys are transferred to the same Reducer. The Reduce phase aggregates the data. In the example, this aggregation is the summation of the values associated with each key, which results in the number of occurrences of each word in the original input. Finally, the outputs of each Reducer are combined to generate the ultimate result.

\subsection{Spark}
MapReduce jobs can be really complex and require many linked Map and Reduce tasks, so many files will have to be generated at runtime. As these files are generated on disk, processing becomes I/O intensive and the cost of complex operations becomes high enough that processing performance is degraded. Spark was created to solve this performance degradation problem in MapReduce, replacing disk structures with distributed memory structures, and eliminating the I/O cost associated with multiple disk reads and writes.

The memory structures were called Resilient Distributed Datasets in the seminal article \cite{zaharia2012resilient}, but today they are called just Dataframes. As we can see in Figure 2, the intermediary files generated by Map and Reduce tasks are placed in memory.

\begin{figure}[ht]
\centering
\includegraphics[width=.60\textwidth]{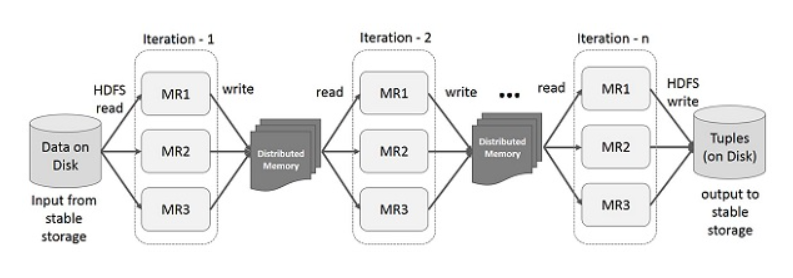}
\caption{Apache Spark Resilient Distributed Datasets} \cite{spark}
\label{fig:exampleFig1}
\end{figure}

Therefore, the conventional method of updating files on disk has been replaced in Spark with in-memory updates distributed across the nodes of the cluster. This feature enables the Map and Reduce tasks to directly access the intermediate processing data in the memory of the cluster nodes, leading to a significant improvement in the performance of the previously inefficient MapReduce jobs. Besides the superior utilization of memory for processing operations instead of disk, one of the distinctive attributes of Spark is its capability to execute interactive applications, such as SQL queries. Spark is considered an extension of MapReduce, which is confined to batch applications, and it provides support for interactive applications including Machine Learning, Streaming, SQL queries, and Graph processing.

\subsection{Polyglot Data Processsing}
The Hadoop ecosystem provides a versatile range of tools for data persistence and processing. Once data is ingested into HDFS, users have the choice to either read the data directly from HDFS or opt for a more specialized storage engine like Hive or HBase, depending on their specific needs. For processing, tools such as MapReduce, Spark, Hive, HBase, and GraphX are available, among many others. 

Hive serves as a data warehousing solution that allows SQL-like querying, converting SQL queries into MapReduce jobs for processing on Hadoop data stored in HDFS, and offers two types of tables for storing data: internal (managed) tables and external tables, each serving distinct purposes and having different implications for data management. HBase, a NoSQL database utilizing a column-family model and running atop Hadoop, offers a specialized storage solution for real-time data processing and querying of large-scale datasets. GraphX, as part of the Spark ecosystem, is specifically designed for effective large-scale graph processing and seamlessly integrates with other Spark components, enabling diverse data analysis tasks.

Polyglot Data Processing in this context involves using various data stores and processing technologies suited to different data types, from structured data in relational databases to unstructured data in NoSQL systems. It combines multiple programming languages and tools for diverse data management, all centralized around HDFS. This approach allows for efficient and scalable data processing, especially in complex areas like social network analysis. Moreover, it provides the flexibility to use specific storage solutions like HBase for certain tasks, thereby optimizing the data processing pipeline for both depth of analysis and efficiency.

\section{Related Work}
Over recent years, numerous researches have been undertaken in the processing of social network data through the utilization of the Hadoop ecosystem, with the goal of gathering, storing, and analyzing data to perform tasks such as sentiment analysis and misinformation investigation. By contrast, research in the field of polyglot data processing using the Hadoop ecosystem is an emerging area that addresses the complexities and challenges of managing diverse data types in large-scale environments and harnessing the power of multiple programming languages, computational paradigms, frameworks and tools within the Hadoop-Spark ecosystem to optimize data processing tasks. 

In \cite{glake2022towards}, the authors explores the concept of polyglot persistence, an approach increasingly recognized for its effectiveness in data management. As detailed in their work, polyglot persistence is strategically designed to combine the advantages of various data stores while avoiding their respective limitations. The article offers a comprehensive overview of polyglot persistence tools, such as Polybase, and delves into related systems like Apache Calcite, providing a thorough summary of these advanced data management solutions.

Several works use the MapReduce programming model to implement social media data analysis. \cite{sheela2016review} proposes the use of Hadoop for processing Twitter data and MapReduce for sentence analysis, text mining and multi-label classification. \cite{sehgal2016sentiment} uses the Twitter streaming API to collect data and MapReduce to perform sentiment analysis over the collected data. In \cite{nandimath2013big}, the authors posit that utilizing JSON files with Hadoop offers benefits in that information is stored in a key-value format, which in turn is used as input by MapReduce.  

As \cite{yang2017design} states, the design of data structure for social network analysis should be based on Hadoop massive datasets interface to meet the requirements of data processing under distributed development environment. The authors use MapReduce for raw data processing and iterative calculation of PageRank value. And in a comparison of data processing tools in Hadoop, such as MapReduce, Hive and Pig, \cite{sachdeva2016comparison} conclude that, when it comes to unstructured data, MapReduce proves to be the most efficient tool.

The prevalent approach employed in the related studies to accomplish their objectives is the utilization of the MapReduce paradigm. However, there exists a noticeable gap within the existing body of literature concerning the utilization of Apache Spark, an alternative distributed computing framework,  conceived as a response to the performance limitations inherent in the MapReduce paradigm and holding particular significance in the domain of big data analytics, particularly when dealing with large-scale datasets such as those found in social networks. 

According to \cite{aziz2018real}, the adoption of in-memory buffers as a replacement for intermediate disk files is what contributes to Spark's superior speed in comparison to Hadoop MapReduce. Indeed, Spark is particularly well-suited for processing large datasets. Moreover, \cite{garg2019sentiment} state that Spark mitigates the need for frequent read and write cycles, resulting in a tenfold improvement in performance compared to Hadoop when processing applications on disk. Additionally, the retention of intermediate data in memory renders Spark a hundredfold faster in memory-intensive scenarios.

Beyond the foundational works associated with the Hadoop ecosystem and the integration of Apache Spark, an emerging field of research is developing, centered on the concept of polyglot data processing. This area of study is dedicated to exploring the integration of multiple processing and storage components, aiming to provide efficient big data solutions, particularly for complex tasks like social network analysis. In parallel, the field is also investigating the strategic use of different storage systems, ranging from HDFS for handling massive datasets to NoSQL databases like HBase, which offer more agile management of unstructured and semi-structured data.

\section{Methodology}
In our current research, the method employed includes a tasks selection phase and a cycle composed by three key steps: ingestion, storage and evaluation. For each task, we adopt a polyglot data processing approach, determining the most suitable processing engines, such as Spark, Hive or MapReduce, and storage strategies, such as HDFS, Hive internal tables or HBase, based on the unique attributes of the job and the nature of the data. This approach allows us to leverage the strengths of various technologies within the Hadoop-Spark ecosystem, ensuring efficient and effective data handling for the selected analytical tasks.

The selected tasks to be evaluated are common inquiries raised by organizations when faced with the challenges of managing and analyzing data originating from social networks. By aligning our research objectives with these real-world concerns, we aim to address the practical needs and expectations of organizations in the domain of social media analytics.

\begin{figure}[ht]
\centering
\includegraphics[width=.50\textwidth]
{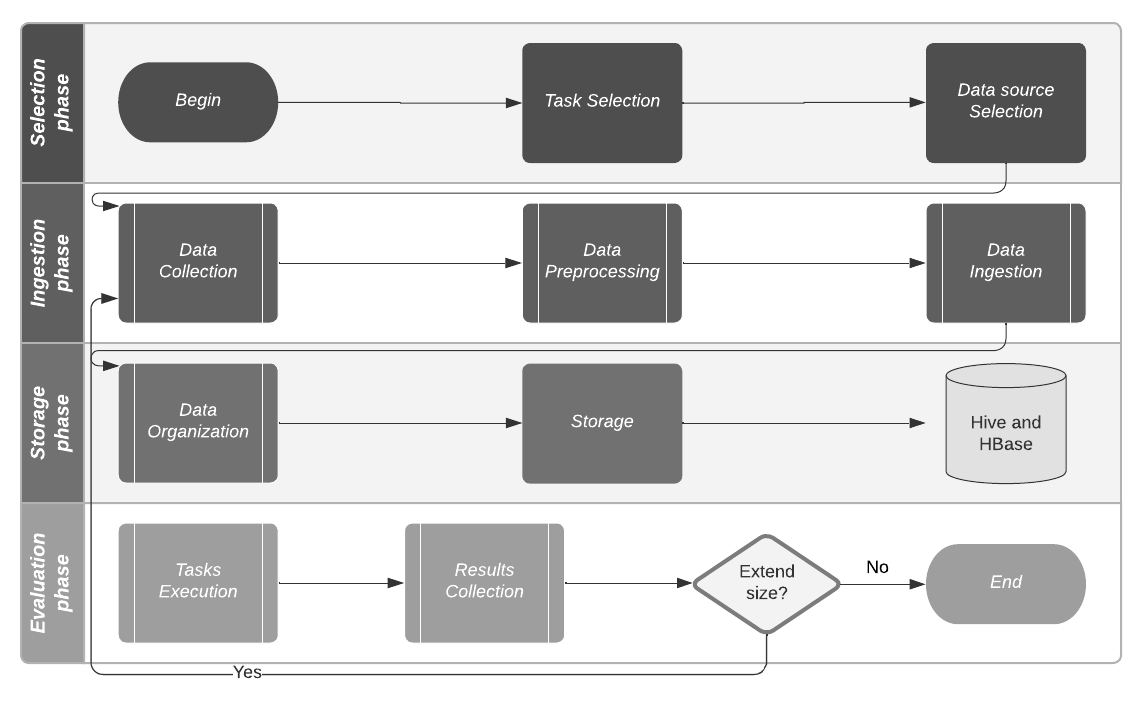}
\caption{Evaluation cycle}
\label{fig:exampleFig1}
\end{figure}

\subsection{Selected Tasks}
In this research, we use data sourced from the Twitter social network. The first task involves determining the most influential users within a specified scope. For each user, we calculate their influence through the number of impressions, likes, quotes, replies and retweets recorded for each posted tweet.

The second task in our evaluation focuses on determining the frequency of terms found in tweets related to a specific entity or topic. To accomplish this, for each tweet we record  all the words used, excluding any stopwords, in order to have a list of the most used words or terms at the end of the processing. 

The third task in our study focuses on examining the relationships between users within the network, aiming to uncover and interpret the various connections existing among users, as well as the affiliations between individual users and groups. 

\subsection{Ingestion phase}
Data ingestion refers to the process of loading data from a source into a centralized repository, where it is stored in its raw format for future processing. In our case, the centralized repository is HDFS and data is collected through the use of the Twitter API, which provides programmatic access to a wealth of diverse and real-time information available on the Twitter platform. 

Data is transferred to a landing zone in HDFS, which plays a pivotal role as it serves as a  staging area where data initially arrives. This landing zone acts as a buffer, allowing for initial inspection and categorization of the incoming data. Following this, the data undergoes a preprocessing stage, in which necessary transformations are applied to ensure data quality, including tasks like cleansing, deduplication, format normalization, and metadata enrichment. Ensuring the quality of data at this stage is vital as it directly impacts the efficacy of subsequent data analytics and processing tasks.

For the purpose of this study, two datasets comprising approximately 500,000 (five hundred thousand) and 5,000,000 (five million) tweets pertinent to the \textit{Last of Us} series were collected. Subsequently, in each cycle, the dataset was preprocessed and organized into two primary files, designated as "\textit{tweets}" and "\textit{users}," for streamlined processing and analysis.

\subsection{Storage phase}
In the storage phase of the polyglot data processing pipeline, a strategic approach is employed to store data across multiple systems beyond the HDFS filesystem, specifically utilizing Hive and HBase. As mentioned earlier, Hive is utilized for its robust data warehousing capabilities, while HBase's columnar storage architecture significantly enhances the efficiency of large-scale data queries by enabling rapid access and retrieval of specific columns within large datasets, thereby optimizing performance for read-intensive operations.

\begin{figure}[ht]
\centering
\includegraphics[width=.70\textwidth]
{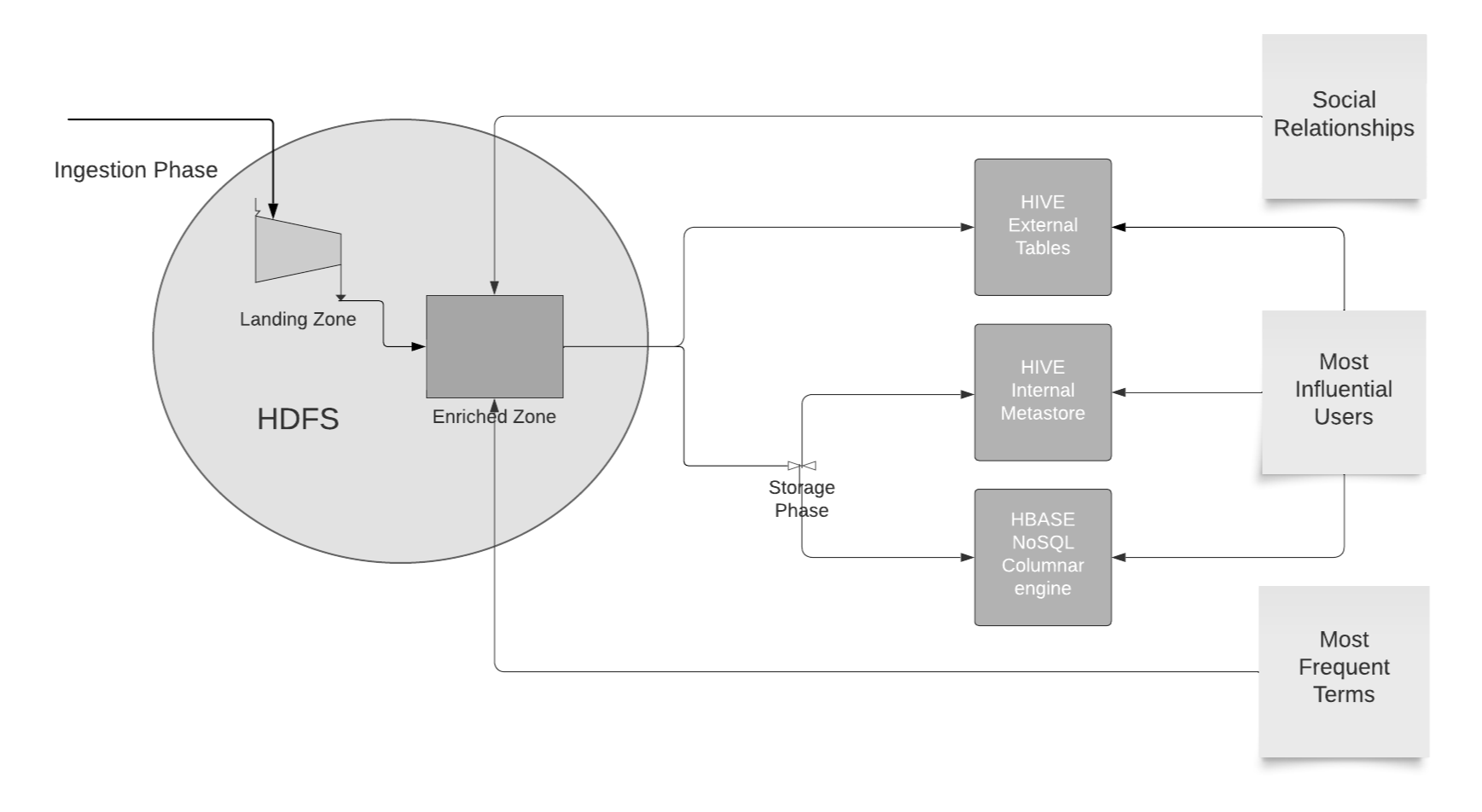}
\caption{Polyglot Architecture}
\label{fig:exampleFig1}
\end{figure}

In order to leverage the capabilities of Hive and HBase storage engines, the preprocessed data must be imported into their respective structures, enabling seamless integration and use within these frameworks. The decision to store the \textit{tweets} and \textit{users} datasets in either Hive or HBase storage engines is strategically determined by the specific computing tasks we aim to evaluate. For instance, a task akin to data warehousing should be assessed using both Hive and HBase, whereas a compute-intensive task would be evaluated through both MapReduce and Spark implementations, each utilizing the same data stored in HDFS.

\section{Evaluation}
Our tasks were all conducted on a Hadoop-Spark cluster implemented over 3 nodes at Google Dataproc service, with each node equipped with 4 cores, 32 GBytes of memory, and a 500 Gbytes SSD disk. This experimental setup used the latest versions of the Hadoop ecosystem components, including Hadoop 3.3.1, HDFS 3.3.0, MapReduce 3.3.1, Spark 3.2.0, Hive 3.1.2, and HBase 2.4.7. 

\subsection{Task 1}
The primary objective of this task is to identify the most influential users within a predefined scope. To achieve this objective, one can create a Hive external table linked to HDFS, a Hive internal table or an HBase table. For the purpose of this task, we simultaneously store tweets data in both a Hive external table and an HBase table. We then execute the same query over the two engines. For each one, we measure and record its execution time. 

By executing these queries and capturing their execution times, we gain valuable insights into the performance and efficiency of the data retrieval and processing operations in each of the mechanism. The query itself, which implements the task, is shown below.

\begin{small}
\begin{verbatim}
SELECT author_id.
SUM(public_metrics.impression_count) AS impressions,
SUM(public_metrics.like_count) as likes,
SUM(public_metrics.quote_count) as quotes, 
SUM(public_metrics.reply_count) as replies,
SUM(public_metrics.retweet_count) AS retweets,
SUM(public_metrics.impression_count) + 
SUM(public_metrics.like_count) + 
SUM(public_metrics.like_count) + 
SUM(public_metrics.reply_count) +
SUM(public_metrics.retweet_count) AS influence
FROM tweets GROUP BY author_id ORDER BY influence DESC
\end{verbatim}
\end{small}

In relation to the HBase data store and the specific task at hand, a column family was established to organize all requisite data for the computation of user influence. This empirical investigation facilitates the comparison of execution times in Hive and HBase, thereby offering valuable insights into the respective processing efficiencies and performance characteristics of these two distinct storage and querying solutions within the Hadoop ecosystem.

\subsection{Task 2}
This task's goal is to determine the dominant terms associated with a specific topic by conducting an iterative analysis of the tweets. To achieve this objetive, we apply both  MapReduce and Spark transformations approaches to systematically iterate through all the tweets, tallying the frequency of terms encountered within the dataset. Both computations use only data stored in HDFS.

\begin{small}
\begin{verbatim}
contadorPalavras = txt.map(lambda x:x.replace(',',' ').
replace('.',' ').replace('-','').lower()) \
.flatMap(lambda x: x.split()) \
.filter(lambda x: x not in stopWords) \
.map(lambda x: (x, 1)) \
.reduceByKey(lambda x,y:x+y) \
.map(lambda x:(x[1],x[0])) \
.sortByKey(False)
\end{verbatim}
\end{small}

The code above is triggered from a Python script running inside a Spark session. By combining the power of Spark with the analysis of the tweet data, we strive to uncover and quantify the most frequent terms pertaining to the chosen topic, facilitating a comprehensive understanding of the linguistic patterns and emphasis within the social discourse.

\subsection{Task 3}
The primary aim of this task is to construct a graphical representation that effectively captures the interconnections among users. By utilizing a table consisting of tuples, wherein each tuple denotes a user \textit{src} following a user \textit{dst}, we can generate a graph that facilitates the analysis and understanding of the communities that have emerged within the user network. 

\begin{small}
\begin{verbatim}
from graphframes import *
df = spark.sql("SELECT id, username FROM users")
df_follows = spark.read.format("csv")
    .option("header", "true") \
    .load("/home/cloud-dataproc/spark-hbase/follows.csv")
g = GraphFrame(df, df_follows)
\end{verbatim}
\end{small}

This graph serves as a valuable tool for extracting insights and discerning patterns in the formation of these communities. All the graphs are generated using the Spark GraphX library and GraphFrames extension, using only data stored in HDFS. 

\subsection{Results}

\begin{wrapfigure}{r}{0.40\textwidth}
\includegraphics[width=0.9\linewidth]
{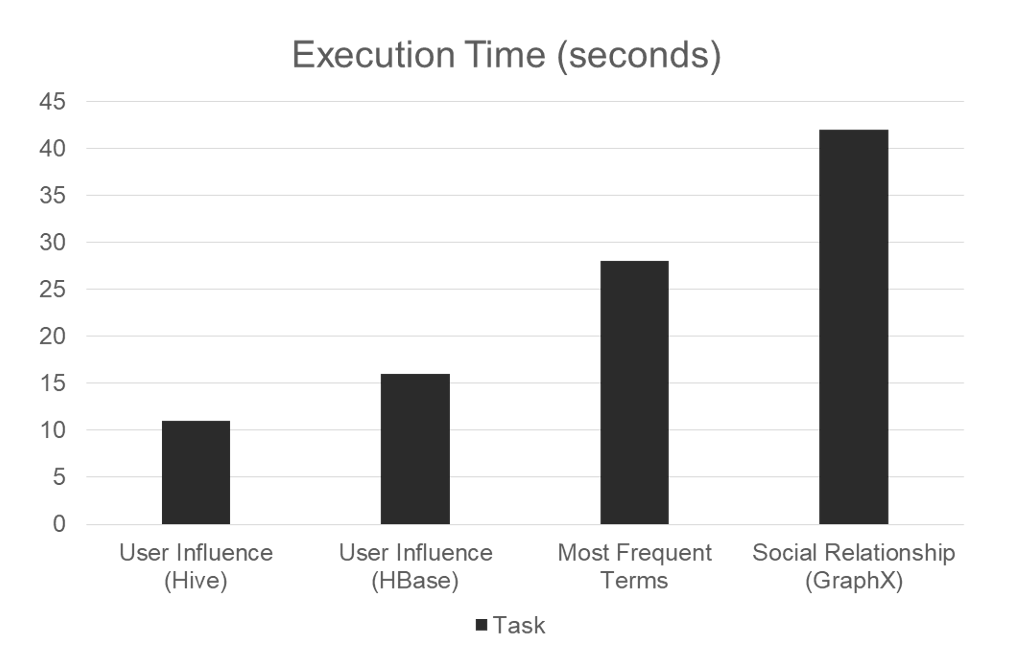}
\caption{Execution times for the 500K dataset} 
\label{fig:exampleFig1}
\end{wrapfigure}

As outlined within the methodological framework, our approach involves a systematic iteration through the stages of ingestion, storage, and processing. We begin this iterative process with the 500,000 tweets dataset. Each job is executed ten times, and the execution times reported in this section represent the average of all the executed trials.

\begin{wrapfigure}{r}{0.40\textwidth}
\includegraphics[width=0.9\linewidth]
{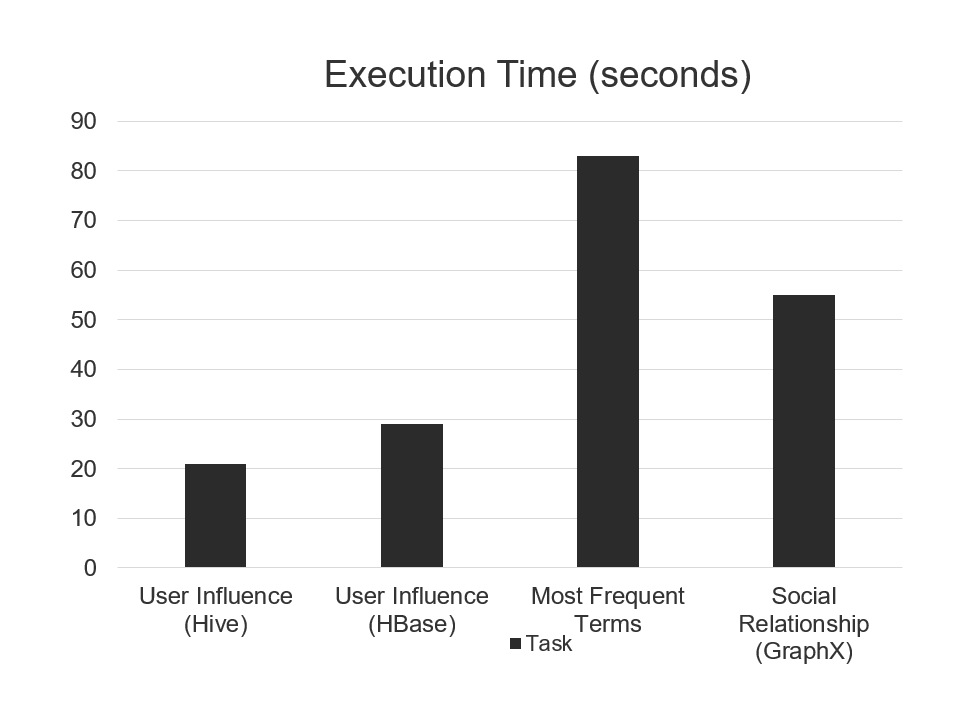}
\caption{Execution times for the 5M dataset} 
\label{fig:exampleFig1}
\end{wrapfigure}

Regarding the first task, our findings indicate that executing a query over a Hive metastore requires approximately 11 seconds, whereas performing the same query over an HBase database takes approximately 16 seconds. For the second task, the MapReduce job implemented for tallying the most frequent terms encountered in the posts exhibits an execution time of approximately 28 seconds, while the proposed graph generation process requires approximately 42 seconds. The time to ingest and store this dataset was less than one minute.

The second dataset utilized in this study consisted of a sizeable collection of 5,000,000 (five million) tweets. The process of ingesting and storing this dataset was executed in an approximate duration of 7 minutes. The obtained results, displayed in the above-mentioned findings, reveal execution times that, remarkably, are notably lower than initially expected, considering the scale of the dataset under investigation.\\

These findings highlight the issue of scalability. The observation that execution times decrease as the dataset volume increases suggests that the solution is likely scalable, warranting further investigation into its capacity to efficiently handle larger datasets.

\section{Conclusions and Future Work}
In conclusion, this research demonstrates the successful utilization of the Hadoop-Spark ecosystem for the processing of data extracted from social networks. The Polyglot Data Processing approach enables us to choose the most appropriate compute and storage engine to process a particular task. Tests executed within the Hadoop-Spark cluster have demonstrated the feasibility of this approach, showcasing the effective use of diverse engines such as MapReduce, Spark, Hive, HBase, and GraphX. Each of these engines offers unique strengths and capabilities, allowing us to tailor our processing strategies to the specific requirements of different tasks. This flexibility not only enhances performance but also optimizes resource utilization, thereby affirming the practicality and efficiency of employing a polyglot methodology in large-scale data processing like the ones we find in social networks challenges.

The methodology implemented in this research is designed to be cyclical, enabling the gradual augmentation of the dataset size, thereby facilitating the assessment of the solution's scalability. A prospective direction for further investigation involves incrementing the volume of collected tweets and subsequently reiterating the tests. This approach would entail cycling through the predefined three stages, meticulously measuring execution times and performance metrics to yield a comprehensive understanding of the system's scalability and efficiency.

Moreover, exploring the configuration of the Hadoop-Spark cluster is crucial. Fine-tuning the cluster parameters, such as the number of nodes, memory allocation, and parallelism settings, can significantly impact the overall performance and scalability of the system. Additionally, exploring query optimization strategies in Hive and HBase, implementing advanced algorithms for data transformation and analysis, and leveraging Spark's built-in optimizations, such as predicate pushdown and data partitioning. Furthermore, the utilization of data compression algorithms can be explored as a means to enhance both storage and processing efficiency. 

Finally, the Hadoop-Spark ecosystem continues to play a pivotal role, particularly with the emergence of new tools and frameworks designed for more efficient and insightful data processing. Notably, Apache Flink emerges as a powerful alternative to traditional batch processing, offering high-throughput, low-latency streaming data processing, ideal for real-time social media analytics. Apache Kafka has become indispensable for managing high-volume data streams from social networks, ensuring robust data ingestion in a distributed environment. Another framework gaining traction is Apache Hudi, which brings the capability of managing storage of large datasets on HDFS, enabling incremental data processing and providing faster access to large-scale social network data. These tools, along with advancements in machine learning libraries like MLlib and deep learning frameworks like TensorFlow or PyTorch integrated with Spark, empower analysts to uncover deeper insights into social behaviors and trends. 

\bibliographystyle{sbc}
\bibliography{social_networks}

\end{document}